\documentclass[12 pt]{article}
\usepackage{natbib}
\usepackage{mathptmx}
\usepackage{amsmath,amstext,amsgen,amsbsy,amsopn,amsfonts,graphicx,theorem,lineno,setspace,graphics,multirow,rotating,lscape,subfig}
\usepackage{setspace}
\begin{document}
%\raggedright
\parindent=0.5in
\pagestyle{myheadings}

%%%%%%%%%
\begin{titlepage}
\begin{center}
{\large \bf Conflict between dynamical and evolutionary stability in simple ecosystems}\\
\singlespacing
Jarad P. Mellard$^{\ast}$ (email: jarad.mellard@ecoex-moulis.cnrs.fr)\\ Ford Ballantyne IV$^\dag$ (email: fb4@uga.edu)\\
\singlespacing
$^\ast$ Centre for Biodiversity Theory and Modelling\\Station d'Ecologie Exp\'erimentale du CNRS\\ 09200 Moulis, France\\ 
$^\dag$ Odum School of Ecology\\
University of Georgia\\		
Athens, GA 30602\\
\singlespacing
%This manuscript is to be published in Theoretical Ecology, 
DOI 10.1007/s12080-014-0217-9\\
The final publication is available at link.Springer.com/article/10.1007/s12080-014-0217-9\\

%{\bf Statement of authorship:}
%JM designed the study and JM and FB performed analyses.  JM wrote the first draft of the manuscript and both authors contributed substantially to revisions.
\end{center}

\begin{abstract}
\begin{normalsize}
Here, we address the essential question of whether, in the context of evolving populations, ecosystems attain properties that enable persistence of the ecosystem itself.  We use a simple ecosystem model describing resource, producer, and consumer dynamics to analyze how evolution affects dynamical stability properties of the ecosystem.  In particular, we compare resilience of the entire system after allowing the producer and consumer populations to evolve to their evolutionarily stable strategy (ESS), to the maximum attainable resilience.  We find a substantial reduction in ecosystem resilience when producers and consumers are allowed to evolve compared to the maximal attainable resilience.  This study illustrates the inherent difference and possible conflict between maximizing individual-level fitness and maximizing resilience of entire ecosystems. 

\end{normalsize}
\end{abstract}
%\newpage
%\noindent

\end{titlepage}
%%%%%%%%%%%%%%%%%%
%\doublespacing

\section{Introduction}

Ecological systems are characterized by processes occurring at different levels of organization, yielding a complicated network of interactions \citep{allenhoek:1992}.  Ecologists have dedicated much effort to understanding how dynamics originating at different levels of organization can feed back upon and constrain one another.  Although the influence of population and community level dynamics on whole ecosystem properties have been documented, for example species richness-productivity relationships \citep{tilman2001}, shifting production to biomass ratios over time \citep{odum1969}, the incidence of N fixers and N availability over succession \citep{vitousek2002}, and diversification influencing primary production and abiotic ecosystem characteristics \citep{harmon2009}, the degree to which evolution shapes whole ecosystem properties that subsequently feedback to constrain evolutionary trajectories of constituent populations/species is not known in general \citep{schoener2011}.  Traits at the single species population level, change in response to certain ecological pressures (e.g. resource acquisition, predator defense), but it is not clear if evolutionary optima for individual species beget the greatest resilience to perturbation at the ecosystem level.  

Characterizing how systems respond to perturbations is, at its core, an analysis of stability, and a
 hallmark of theoretical studies in ecology \citep{may:1973}.   From analyzing a variety of multi-species models, we have learned that species richness and food web structure may alter the stability of communities \citep{may:1973,mccanetal:1998,Loreau1998,Thebault2005,Vallina2011}, but in most theoretical studies, model parameters are typically chosen to maintain coexistence of all the species in the system rather than allowing trait combinations across species to emerge through mutation, speciation, and extinction.  In real ecosystems, component species are subject to ecological pressures arising from interactions between conspecifics and other species, which determine selection pressures that may subsequently affect dynamical stability properties of the ecosystem.  Behavioral changes, species replacements, and changing allele frequencies all have the potential to alter the nature of interspecific interactions, which by influencing rates of resource use and predation, can radically shift trophic structure \citep{estes1998,harmon2009} and ecosystem processes \citep{walsh:carbon_budget,miner2012,bassar2010}.

Theoretical work on evolution to ecological attractors has blossomed over the past 20 years in Ecology.  Earlier during this period, \citet{marrow1996} called for addressing whether general mechanisms of selection lead to different types of dynamics  because some studies showed that evolution leads to chaos \citep{ferriere1993} while others showed that populations evolve to simple dynamics \citep{doebeli1995}, and some even demonstrated that evolution may cause unstable (ecologically) systems to become stable \citep{hochberg1995}.  When the evolution of traits has been explicitly incorporated into ecological models, adaptive behavior along fitness gradients can confer desirable whole-system stability properties \citep{Valdovinos2010, loeuille2010,doebeli:monograph}, destabilize dynamics \citep{Loreau2010}, or have negligible effects on local stability \citep{lawlor1976}.  Examining the first directional step in evolution via ecological selection, \citet{loeuille2010} showed that evolution in an ecological context can increase or decrease dynamical stability for relatively generic models, and that the ecological context (interaction type and diversity) is critical for determining community stability.  Using evolution to guide assembly, uninvasible (evolutionarily stable) communities can be constructed, but if they exist in regions of dynamic instability, species can go extinct due to large oscillations \citep{wilson1996}.  Thus, a point in trait space that is on the cusp of dynamical stability, even if it corresponds to an Evolutionarily Stable Strategy (ESS), is still at peril because a seemingly small mutation can potentially push a system over the precipice.  A well known example is that an increase in consumer efficiency initially results in increased consumer fitness, but can lead to species extinction \citep{Webb2003,Loreau2010}(Ch. 4), analogous to the paradox of enrichment \citep{rosenzweig1971}.

Many studies have assumed that ecosystems organize in response to fundamental principles, often maximizing or minimizing a property or flux \citep{jorgenson:thermoecology}.  \citet{schneider:kay} propose that ecosystems evolve to maximize energy dissipation, \citet{mulleretal:indicators} argue that ecosystems maximize integrity, \citet{cropp2002} suggest that resilience is maximized, \citet{odum1969} hypothesizes that production to biomass ratios are minimized at equilibrium, and \citet{pattee:simplification} takes an alternative view that living systems self-simplify.  A unifying theme across all of these studies is some degree of self-reinforcement \citep{ulanowicz1997} of core ecosystem structure.  In a thought-provoking study, \citet{cropp2002} examined properties of ecosystems that maximized certain ecological goal functions, such as resilience and a number of related thermodynamic properties, and found that high maximum growth rate of the autotrophs, low grazing/palatability, and high herbivore excretion led to high ecosystem resilience.  However, this means that the herbivores in the ecosystem were very inefficient and hence poor competitors, and is at odds with the fact that the most competitive herbivores should reduce the autotrophs to the lowest level \citep{holt1977}.  

Here, we aim to address the essential question of whether in the context of evolving populations, ecosystems attain properties that promote persistence of the ecosystem itself.  We use a coupled resource-producer-consumer model similar to \citet{cropp2002, loeuille2002, loeuille2004} to analyze how evolution affects whole ecosystem resilience.  Rather than making resilience or some other whole-system properties a goal for evolution, we focus on the end result of evolution in terms of dynamical stability properties.  We allow the evolution of a producer species subject to a tradeoff between nutrient acquisition and grazing from consumer species, and the evolution of a consumer species subject to a tradeoff between resource consumption and mortality.  When only the producer species evolves, we show how the location of the ESS in nutrient acquisition and grazing-rate trait space changes along an environmental gradient in consumer strategies.  When only the consumer species evolves, we show how the location of the ESS in grazing-rate and mortality trait space changes along an environmental gradient in producer strategies.  When both the producer and consumer simultaneously evolve, we quantify the different potential coevolutionary outcomes.  For each evolutionary scenario, we characterize how the entire ecosystem responds to perturbations by quantifying whole system resilience, and we compare the resilience of the ESSs to the maximum resilience.  We establish that  optimal traits for population-level fitness often do not correspond to those that maximize the resilience of a generic and widely-used ecosystem model.

\section{Material and methods}

\subsection{Model}

We build on previous modeling of food chains that are subject to dynamical \citep{cropp2002} and evolutionary \citep{loeuille2002, loeuille2004} selective forces.  We define a simple ecosystem with equations for the inorganic nutrient resource $R$, producer species $P$, and consumer (herbivore) species $H$,
\begin{eqnarray}
\frac{dR}{dt} &=& I - R (q+k P)\\ 
\frac{dP}{dt} &=& P(k l R - m - a H)\\ 
\frac{dH}{dt} &=& H(-d+ a b P), \label{popeqns}
\end{eqnarray}
where $I$ represents inorganic nutrient input, $q$ represents the inorganic nutrient loss rate, $k$ is the nutrient uptake rate, $l$ is the conversion factor of nutrients into producers, $m$ is the mortality rate of producer species $P$, 
$a$ is the consumption rate of the consumer species on the producer species, $b$ is the conversion factor of the producer species into the consumer species, and $d$ is the mortality rate of consumer species $H$, similar in notation to \citet{hulot2006}, see also Table 1.

For our analysis, let $\hat{R}, \hat{P}, \hat{H}$ represent equilibrium densities so that,
\begin{eqnarray}
\hat{R} &=& \frac{I}{q+\Omega\Psi}, \\
\hat{P} &=& \frac{\Psi}{a}, \\
\hat{H} &=& \frac{I l}{\frac{q}{\Omega}+\Psi}-\frac{m}{a},
\end{eqnarray}
is the equilibrium with the producer and the consumer present.  The existence of this interior equilibrium with $\hat{H}>0$ is possible when $I > \Psi \frac{m}{a l} + \frac{m q}{k l}$. 
We have made the following substitutions for simplification:  
\begin{equation}
\Omega=\frac{k}{a},
\quad \quad
\Psi=\frac{d}{b}.  
\end{equation}
%%%%%%%%%%
$\Omega$ is a ratio of per capita per exploiter loss rates, and thus reflects the efficiency with which producers and consumers deplete or take up lower trophic level resources.  Because $\Omega$ is inversely related to steady state nutrient concentration and positively associated with consumer biomass, it integrates resource fluxes across two trophic levels in a way that neither the standard conversion factor $b$ nor producer uptake do.  $\Psi$ is the effective per capita loss rate from the consumer.  Ecosystems with high $\Psi$, therefore, do not effectively convert producers to consumer biomass and retain the converted biomass in the consumer pool, rendering a greater fraction of the initially consumed basal resources underutilized.  $\Omega$ is a very useful quantity being a ratio of parameters that define the ecological interactions in the simple ecosystem that are ultimately determined by traits of the organisms themselves (see below).  Inspection of the Jacobian matrix evaluated at equilibrium,
\[
\hat{J}=
\begin{pmatrix}
-q-\Omega\Psi & -\frac{I k}{q+\Omega\Psi} & 0 \\
\Omega\Psi l & 0 & -\Psi \\
0 & \frac{I a b l}{\frac{q}{\Omega} + \Psi} -m b & 0 
\end{pmatrix} \label{jacobian}
\]
makes it clear that whole system stability is affected by not a single parameter but a multitude of quantities, with $\Omega$ likely to be important.  The above equilibrium is stable (see Appendix for details), which is to be expected given its similarity to the model used by \citet{loeuille2002}, \citet{loeuille2004} and \citet{cropp2002}.

%%%
\begin{landscape}
\begin{table}[htb] 
\caption{Parameter values unless noted otherwise.}
\centerline{ \begin{tabular}{lll} \hline Variable or Parameter [Dimension] & Definition & Value \\ \hline 
	$ R $ [Quantity of nutrient] & inorganic nutrient concentration & state variable \\
	$ P $ [Quantity of biomass] & producer species biomass & state variable \\
	$ H $ [Quantity of biomass] & consumer (herbivore) species biomass & state variable \\ 
	$ s_P $ [Dimensionless] & producer trait determines nutrient uptake and consumption by consumer & evolves \\
	$ s_H $ [Dimensionless] & consumer trait determines consumption strategy & evolves \\
	\\
	$ I $ [nutrient (time)$^{-1}$] & inorganic nutrient input & 5 \\
 	$ q $ [time$^{-1}$] & inorganic nutrient loss rate & 1 \\
	$ k $ [(producer biomass x time)$^{-1}$] & per capita nutrient uptake rate & function of $ s_P $ \\
	$ l $ [producer biomass (nutrient)$^{-1}$]  & conversion factor (efficiency) of nutrients into producers & 1 \\ 
	$ m $ [time$^{-1}$] & mortality rate of producer species & 0.2 \\ 	
	$ a$ [(consumer biomass x time)$^{-1}$] & per capita consumption rate of the consumer species on the producer species & function of $ s_P,s_H $ \\
	$ b$ [consumer biomass (producer biomass)$^{-1}$] & conversion factor (efficiency) of the producer species into the consumer species & 1\\
	$ d$ [time$^{-1}$] & mortality rate of consumer species & function of $ s_H $ \\ 	
	\hline	
\end{tabular}}
\end{table}
\end{landscape}
%%%

\begin{figure}[h!] 
\centering \leavevmode \includegraphics[scale=0.9]
	{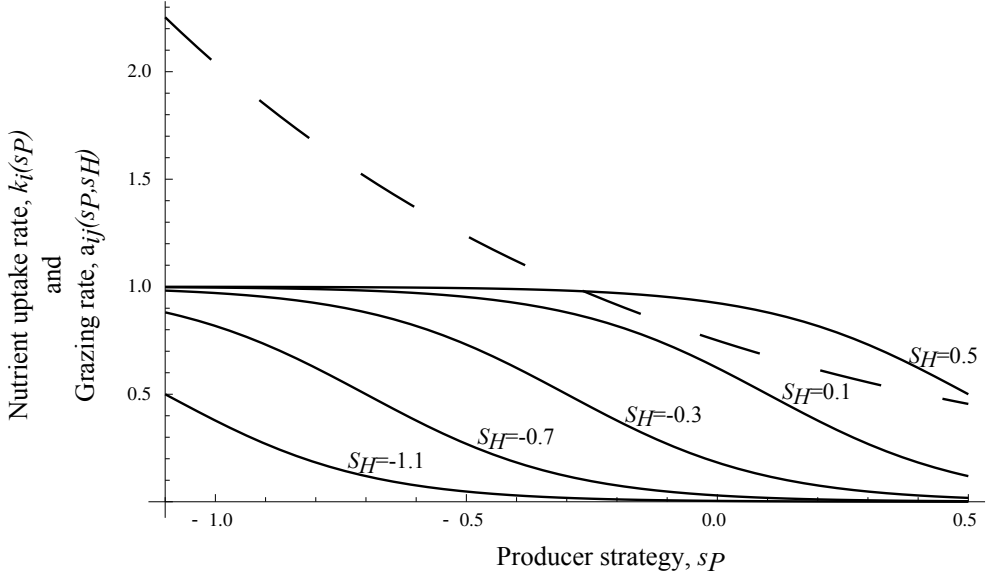}
	 \caption{Nutrient uptake rate [(producer biomass x time)$^{-1}$] and grazing rate [(consumer biomass x time)$^{-1}$] parameters $k(s_P)$ and $a(s_P,s_H)$ as a function of producer strategy, $s_P$.  Dashed line is a plot of  $k(s_P)$ where $\omega=0.75$.  Solid lines are plots of $a(s_P,s_H)$ where $\alpha=1, \beta=1, \chi=5$, for five different values of consumer strategy, $s_H$ (range is -1.1 to 0.5, the line nearest $a(s_P,s_H)=0$ is for $s_H=-1.1$).  
\label{chiplotsallgrazweak}  
	} 
	 \end{figure}

To explore how traits and trade-offs influence ecological interactions, resource uptake, and grazing pressure, we incorporate several aggregate traits for species in the system that define strategies in resource consumption similar to \citet{loeuille2002}.  The producer species, $P$, has a trait $s_P$ that defines its strategy in nutrient uptake and susceptibility to grazing.  The consumer species, $H$, has a trait $s_H$ that defines its strategy in grazing and mortality.  Specifically, we let nutrient uptake ($k$) be a function of $s_P$, so that $k(s_P)=\omega e^{-\gamma s_P}$, see Figure \ref{chiplotsallgrazweak}.  We let $a$ be a function of $s_P,s_H$ to define the grazing interaction so that $a(s_P,s_H)= \frac{\alpha}{\beta +e^{\chi(s_P-s_H)}}$, see Figure \ref{chiplotsallgrazweak}.  Note that low values of $s_P$ have high nutrient uptake rates but are also grazed at the highest rate.  This corresponds to the situation in which herbivores have a preference for more nutritious producers \citep{branco2010}. 
For the consumer species, we assume that mortality rate increases as a function of the trait $s_H$ that increases grazing rate ($d(s_H)=d_H e^{g s_H}$), inducing a tradeoff.  This could occur if consumers experience greater metabolic losses or more intense predation as their own grazing rates increase \citep{loeuille2002}.

\subsection{Adaptive trait evolution}

To study how trait evolution influences whole ecosystem properties and ecological interactions, we borrow many of the tools of adaptive dynamics \citep{Dieckmann1996,geritz1998} to analyze trait evolution.  
We make the standard assumption, that mutations are rare and occur infrequently enough for populations to reach their ecological attractor before a new mutation occurs, therefore creating a separation of ecological and evolutionary time scales.  Furthermore, we assume that mutations result in small magnitude changes in trait values, but that they correspond to discrete steps in phenotypic space.  Typically, reproduction is assumed to be clonal \citep{geritz1998}.

First, we consider phenotypic change due to evolution within the producer species.  We define the fitness, $W_{P_{inv}}$, of an invading phenotype of the producer species with trait $s_{P_{inv}}$

\begin{equation}
W_{P_{inv}}(s_{P_{inv}},s_P,s_H)=\frac{1}{P_{inv}} \frac{dP_{inv}}{dt} = (k(s_{P_{inv}}) l \hat{R} - m - a(s_{P_{inv}},s_H) \hat{H})
\end{equation}

Similarly, we define the fitness of an invading phenotype of the consumer species, $W_{H_{inv}}$ with trait $s_{H_{inv}}$
\begin{equation}
W_{H_{inv}}(s_{H_{inv}},s_P,s_H)=\frac{1}{H_{inv}} \frac{dH_{inv}}{dt} = (-d(s_{H_{inv}}) + a(s_{P_{inv}},s_H) b \hat{P})
\end{equation}
and evaluate both fitness equations with the equilibrium environment set by the resident species $P$ and $H$ (refer to the equilibrium above).  Hereafter we use the following subscript notation to describe: the $j=P$ producer species and $j=H$ for the consumer species evolutionary equations involving derivatives of the fitness equations.

\subsubsection{Finding the singular strategies}

Our goal was to find the evolutionary endpoints for a given abiotic environment.  More specifically, we wanted to determine the trait values species eventually evolve to on the fitness landscape.
We can directly find these trait values by simultaneously setting the equations for the fitness gradients ($\left. \frac{\partial W_{j_{inv}}}{\partial s_{j_{inv}}} \right|_{s_{j_{inv}} \rightarrow s_j} $ for $j=P,H$) equal to zero and solving for the trait values.  These are known as the singular strategies \citep{geritz1998}.  The analytical expressions for singular strategies, while they exist, do not permit interpretation in general because the expression for the producer singular strategy is too complicated, hence we resorted to numerical methods in all producer cases.  Specifically, in Mathematica, we computed derivatives analytically (using ``D") and then parameterized the expressions and used numerical root finding techniques (using ``FindRoot")  to locate the singular strategies.  For the consumer, we can obtain an analytical result by setting 
$\left. \frac{\partial W_{H_{inv}}}{\partial s_{H_{inv}}} \right|_{s_{H_{inv}} \rightarrow s_H} =0$ and solving for $s_H$, which yields the consumer evolutionary isocline, $s_H = s_P + \frac{ \rm{ln} (\frac{\chi-g}{g})}{\chi}$.

To confirm the validity of our numerically determined singular strategies, we routinely performed evolutionary simulations.  The goal of these simulations was two-fold:  1)  to confirm the location of the singular strategy found numerically as described above, and 2) for coevolution, to confirm that runaway selection was the outcome (runaway selection is defined as ever increasing or decreasing trait values) when trait values were unconstrained.  For the simulations, first we defined the dynamics of the traits ($s_P$ and $s_H$ of the producer and consumer respectively) evolving through mutation and selection as
\begin{equation}
\frac{ds_j}{dt}= \mu \hat{N}_j \left. \frac{\partial W_{j_{inv}}}{\partial s_{j_{inv}}} \right|_{s_{j_{inv}} \rightarrow s_j} \; \quad \quad \text{for $j=P,H$}
\label{canonical}
\end{equation}
in which the mutation rate, $\mu$, dictates the pace of evolutionary change and $\hat{N}_j$ is the equilibrium population size for $j=P,H$.  For each simulation, we numerically solved the five-dimensional system consisting of the three equations for the quantities of the resource, producer, and consumer (equations 1, 2, 3) and one equation for each of  the trait values of the producers and consumers (equations \ref{canonical}).  Thus, all five state variables were allowed to change simultaneously.  In the evolutionary simulations with Equations \ref{canonical}, we routinely tested multiple sets of initial conditions and allowed traits to evolve until no further change was observed.
Coevolutionary cycles can occur if parameters are chosen to be within a limited range and rates of evolution between producers and consumers are disparate enough \citep{loeuille2002} .  Our purpose here was not to detail all the qualitatively different evolutionary cases, as this has already been shown in \citet{loeuille2002}, rather we sought to determine the evolutionary endpoints and their corresponding values of resilience and compare them to maximum resilience, thus we set  $\mu$ to be equal between the species.

\subsection{Classifying the singular strategies}

To classify the singular strategies, which we denote as $s_j^*$, we determined the signs of the following derivatives numerically (using the same parameters as simulations).
We take the second derivative of the invader fitness equation with respect to the invader
\begin{equation}
\left. \frac{\partial^2 W_{j_{inv}}}{\partial s_{j_{inv}}^2} \right|_{s_{j_{inv}} \rightarrow s_j \rightarrow s_j^*}\; \quad \quad \text{for $j=P,H$}
\label{ESS}
\end{equation}
and find that Eqn \ref{ESS}$<0$ for $j=P$ or $H$, therefore the singular strategy is an ESS for the producer or consumer species respectively.  To further classify the ESS, we take the second derivative of the fitness equation with respect to the resident 
\begin{equation}
\left. \frac{\partial^2 W_{j_{inv}}}{\partial s_{j}^2} \right|_{s_{j_{inv}} \rightarrow s_j \rightarrow s_j^*} \; \quad \quad \text{for $j=P,H$}
 \label{singularity}
\end{equation}
to show that it is convergence stable (Eqn \ref{singularity}-Eqn \ref{ESS}$>0$ for $j=P$ or $H$) and therefore a continuously stable strategy, CSS (Geritz et al.1998)  for the producer or consumer species respectively.  These findings are consistent with the conclusions of \citet{loeuille2002, loeuille2004} that for a set of parameters in similar models, there can exist an ESS that is convergence stable and it is this CSS on which we concentrate for the three different cases of evolution:  producer evolution only, consumer evolution only, and coevolution.  

\subsection{Does evolutionary stability match maximum resilience?}

 For the model ecosystem considered here, the ecologically relevant equilibrium (both producers and consumers positive) is stable \citep{loeuille2002}, and CSSs for evolving producers and consumers can also exist \citep{loeuille2002}, but our interest was in determining how producer and consumer trait evolution influences resilience of the entire ecosystem.  Although model equilibria are typically classified as stable or unstable in a binary sense \citep{Allesina2012}, different parameter sets can result in quite different dynamical responses following a perturbation, namely the rate of return to equilibrium or resilience.  We quantified relative stability of different stable equilibria using a resilience metric, which characterizes how fast a system returns to equilibrium \citep{Pimm1984}.  Return time is often approximated by the inverse of the absolute value of the real part of the eigenvalue (with the largest real part) because it is highly correlated with return time for an entire system \citet{deangelis1989, cottingham1994}.  Return time depends on the type of perturbation so although they are qualitatively similar, measured return time can differ from the return time computed from the dominant eigenvalue.

We calculated the resilience of the ecosystem described by Eqns 1,2,3 as a function of both producer and consumer traits.  We followed \citet{deangelis1980} and defined resilience of the entire ecosystem  (Eqns 1,2,3) as the absolute value of the real part of the eigenvalue with the largest real part of the Jacobian matrix for the system: \begin{equation} {\rm Resilience}= -{\rm Max[Re}(\lambda)]. \label{resilience} \end{equation}
We compared this measure of resilience to resilience measured as the inverse of return time following a perturbation and found general concordance.  However, we only present results from resilience calculated from eigenvalues for greater tractability and to follow previous studies \citep{cropp2002, loeuille2010}.

We were also interested in what properties of our model determined measured resilience \citep{deangelis1980,deangelis1989,cropp2002}.  Unfortunately, the analytical expressions for the eigenvalues are sufficiently complicated to prevent any inference about how parameters or combinations of parameter such as $\Omega$ influence resilience from simple inspection.  As a consequence, we numerically computed $-{\rm Max[Re}(\lambda)]$ over a range of parameters and trait values of $s_P$ and $s_H$ to quantify resilience.  This allowed us to illustrate the exact influence of combinations of parameters such as $\Omega$ on resilience.  Overall, for results presented, we varied the following parameters (and their range):  $\chi$ (1 - 13), $\gamma$ (0.24 - 0.85), $g$ (0.24 - 0.99), and $a, k, d$ (0 - 1) .

\section{Results}

\subsection{Traits, ecological interactions, and resilience}

\begin{figure}[h!] \centering \leavevmode \includegraphics[scale=0.9]
	{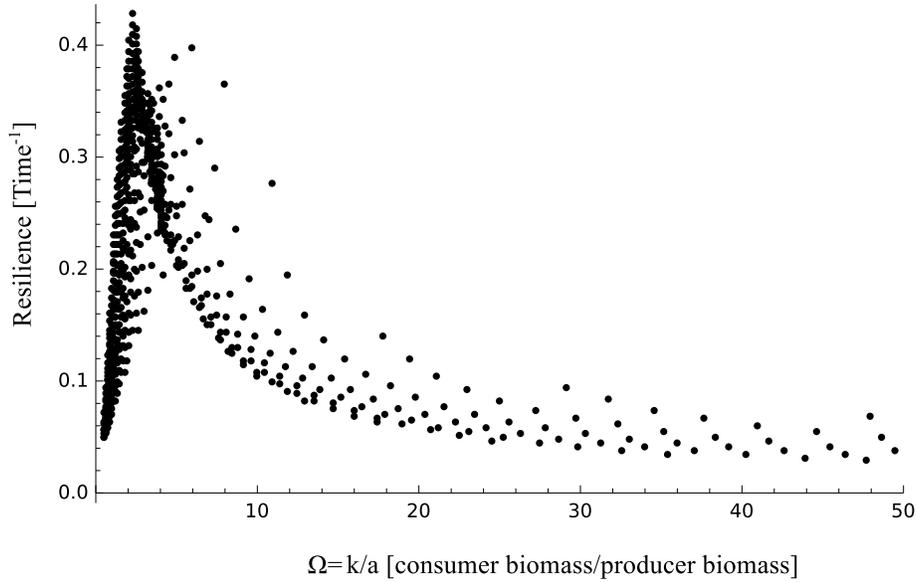}
	 \caption{Resilience for the system as a function of $\Omega$.  We generated this figure by sampling with a regular grid a region of $s_P - s_H$ parameter space that determine resilience, and plotting those values of resilience against the values of $\Omega$ for that grid in the region of $s_P - s_H$ parameter space.  Region of $s_P - s_H$ parameter space is the same as Figure \ref{contourer} except only where $H$ is positive.
Sampled locations create n=699 points.  Parameters are:  $\alpha=1, \beta=1, \chi=5, g=0.95, d_H=0.05, \omega=0.75, \gamma=0.85$.  	 \label{ka}  
	} 
	 \end{figure}

We predicted that the qualitative influence of traits on stability will be related to $\Omega$, because in all but one entry in the Jacobian for the system, $k(s_P)$ and $a(s_P,s_H)$ appear as the ratio $\Omega$, and $k(s_P)$ appears in only one entry without $a(s_P,s_H)$.  Although $d$ is a function of $s_H$, and thus evolution of $s_H$ may influence stability via $\Psi$, we expected $\Omega$ to be more influential because of previous work \citep{cropp2002} and because it is a function of two evolving traits (and thus a richer ratio than $\Psi$).  In Figure \ref{ka}, we see that $\Omega$ indeed accounts for the majority of variation in resilience and that maximum resilience occurs at an intermediate value.  The residual variation in Figure \ref{ka} results from the non-uniqueness of $\Omega$ as a function of $s_P$ and $s_H$.  It is possible for the same $\Omega$ value to be reached with different combinations of $s_P$ and $s_H$, but because $\Psi$ is an independent function of $s_H$, different $s_P$ values result in different levels of resilience even if $\Omega$ does not vary.  This means that ecosystems (with relatively low per capita per exploiter loss rate ratios) with intermediate nutrient concentrations and consumer biomass, confer the greatest resilience for the ecosystem model studied here.  Computing the resilience landscape as a function of traits subject to evolution (Figure \ref{contourer}) sets the stage for determining whether trait evolution moves ecosystems toward regions of increased or decreased resilience.

\subsection{Evolutionary outcomes on the resilience landscape}

In general, the particular outcome of evolution (runaway, node, or cycle) crucially depends on which  traits are evolving (producer and consumer traits in isolation or both simultaneously) and how much the parameters defining ecological interactions and resilience (such as $g, \chi, \gamma$) cause changing trait values \citep{loeuille2002} to impact ecological and evolutionary dynamics.  By directly placing evolutionary attractors (specific trait values) in the context of the resilience landscape, we were able to determine if traits associated with individual ecosystem components evolve to reinforce whole ecosystem stability.  We see in Figure \ref{contourer}a that resilience exhibits a complex relationship with the producer strategy, $s_P$ and consumer strategy, $s_H$.  
Of particular note, for a fixed $s_H$ strategy, resilience can be a multi-modal function of $s_P$.  This leads to multiple local peaks in resilience along a gradient of  $s_P$. 

\subsubsection{Isolated evolution of ecosystem components}

\begin{figure}[h!] \centering \leavevmode \includegraphics[scale=0.7]
	{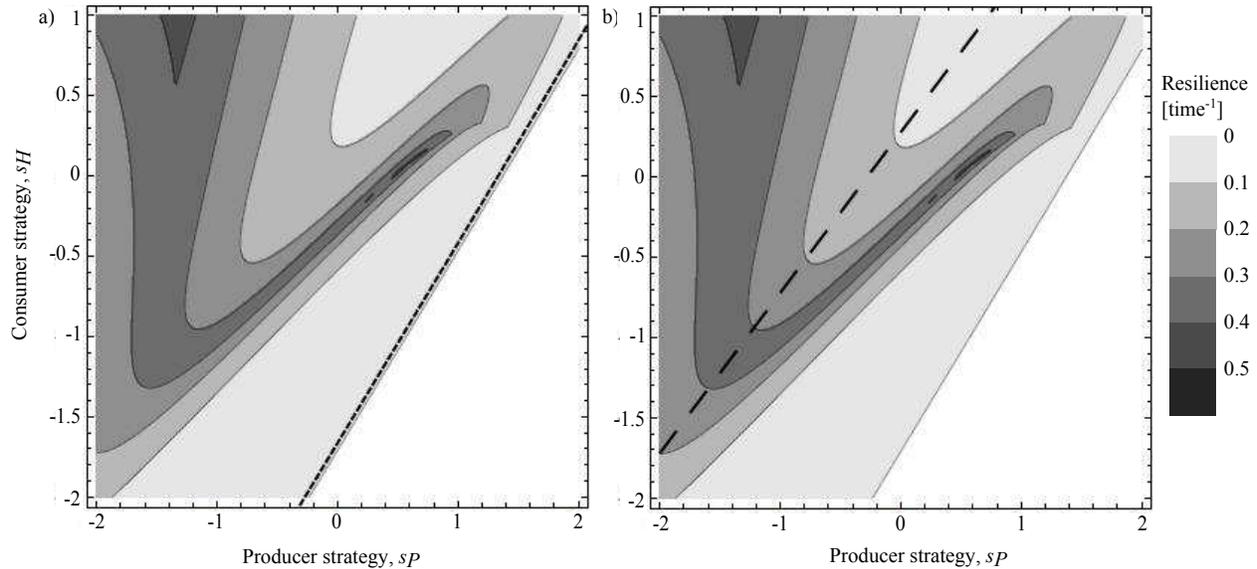}
	 \caption{Resilience as a function of producer strategy, $s_P$ and consumer strategy, $s_H$ for a) producer evolution only and b) consumer evolution only.  Darker regions are higher levels of resilience.  Resilience calculated according to equation \ref{resilience}.  In a), the trait values that comprise the producer CSS are denoted by the small dashed line on the plot and in b),  the trait values that comprise the consumer CSS are denoted by the large dashed line on the plot.  The white region is a feasibility boundary where consumers cannot exist in the system ($\hat{H}=\frac{I l}{\frac{q}{\Omega}+\Psi}-\frac{m}{a} \leq 0$).  Resilience values shown in legend for other shades of gray.  Parameters are:  $\alpha=1, \beta=1, \chi=5, g=0.95, d_H=0.05, \omega=0.75, \gamma=0.85$.
	  \label{contourer}  
	} 
	 \end{figure}

For the model studied here, when only the producer trait is allowed to evolve, the CSS for the producer species in the $s_P$ - $s_H$ trait space traverses a region of low resilience.  In fact, in Figure \ref{contourer}a, the producer CSS is in a region of parameter space that exhibits generally less than $25\%$ of the potential maximum resilience.  Furthermore, the CSS line is very close to the boundary of consumer extinction; the white region in Figure \ref{contourer}a is where consumers cannot exist in the system ($\hat{H}= \frac{I l}{\frac{q}{\Omega}+\Psi}-\frac{m}{a} \leq 0$).  Thus, selection will move the system away from regions of highest resilience to the CSS.  The discrepancy between evolved resilience and maximum resilience apparent in Figure \ref{contourer}a is representative of the parameter space we explored (see appendix).  We compared this result across other parameter values as depicted in Appendix Figure \ref{gridproev} to illustrate robustness of this result.

When only consumers evolve, the situation is similar to isolated producer evolution (Figure \ref{contourer}b), however, note that the consumer CSS isocline ($s_H = s_P + \frac{ \rm{ln} (\frac{\chi-g}{g})}{\chi}$) traverses a small region of trait space exhibiting high resilience, roughly up to $80\%$ of the maximum resilience in Figure \ref{contourer}b.
  By varying $g$ and $\chi$, which influence the steepness of producer and consumer trade-offs, it is possible for the CSS isocline to be closely aligned with the ``left-hand thumb" region of high resilience (for example by increasing $\chi$), but as the CSS approaches the region of increased resilience, the stability landscape changes, the region of high resilience shrinks, and the isocline never intersects with the trait combination yielding maximum resilience (see appendix).

\newpage
\subsubsection{Coevolution}

\begin{figure}[h!] \centering \leavevmode \includegraphics[scale=0.7]
	{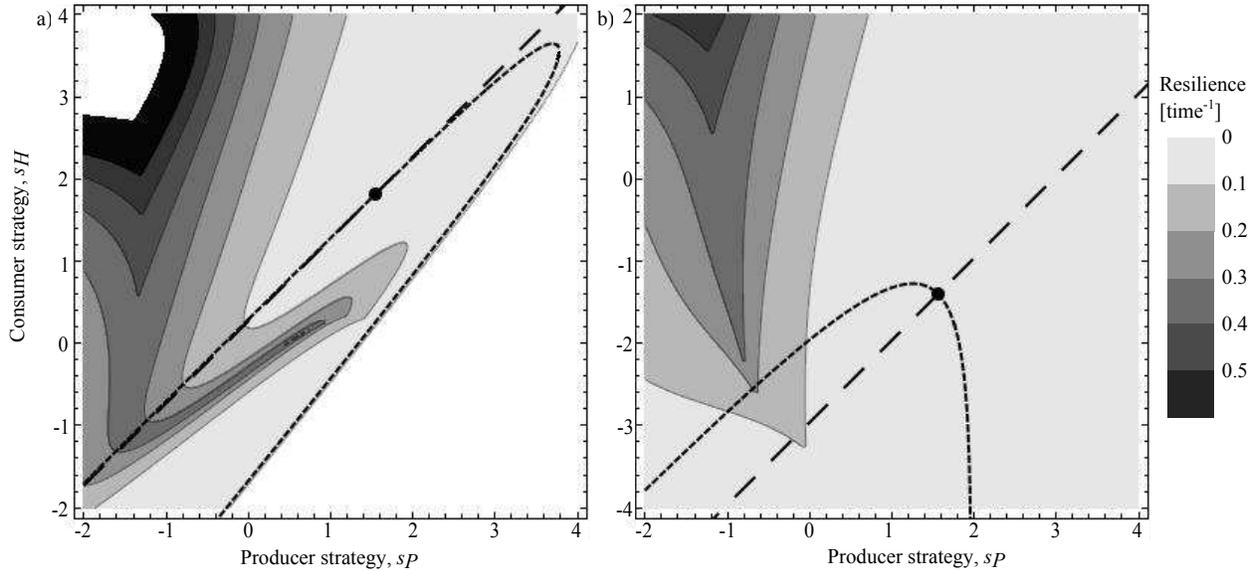}
	 \caption{Resilience as a function of producer strategy, $s_P$ and consumer strategy, $s_H$ for coevolution to a) consumer CSS and b) joint CSS.  Darker regions are higher levels of resilience.  However, for a), the upper left white region is for resilience $>0.6$.  Resilience calculated according to equation \ref{resilience}.  The producer isocline is denoted by the small dashed line and the consumer isocline is denoted by the large dashed line on the plot.  The trait values that comprise the one (a) or two (b) species CSS are denoted by the black dot on the plot.  Resilience values shown in legend for other shades of gray.  Parameters for a) are:  $\alpha=1, \beta=1, \chi=5, g=0.95, d_H=0.05, \omega=0.75, \gamma=0.85$.  Parameters for b) are the same as a) except:  $\chi=1$. 
	  \label{coevcss}  
	} 
	 \end{figure}

When both the producer species and the consumer species are allowed to simultaneously evolve, the possible evolutionary outcomes are increased to at least four cases, depending on the positions of the producer and consumer isoclines  \citep{loeuille2002}.  A CSS for one or both species (joint CSS) is possible.  However, for most parameters, the typical outcome is runaway selection \citep{loeuille2002} which pushes both $s_P$ and $s_H$ to negative infinity since we do not artificially bound the trait values.  Runaway selection does not yield ecosystems with high resilience.
If the consumer isocline intersects the producer isocline left of its maximum (Figure \ref{coevcss}a), the result is instability, but evolution can lead to stabilization if the mutation rate of the producer relative to the consumer is below a certain threshold \citep{loeuille2002}.  However, this case is a CSS only for the consumer, not the producer.  In fact, it is an evolutionary repeller for the producer.  Regardless, the stable evolutionary endpoint is in a region of low resilience (Figure \ref{coevcss}a), generally less than $25\%$ of the potential maximum resilience. 

If the consumer isocline intersects the producer isocline on the stable part, right of the maximum, the result is a CSS for both species, but this joint CSS can be somewhat difficult to achieve because it requires the evolutionary stability of not just one but both species.  Additionally, the region of parameter space that results in a joint CSS appears to be small relative to the large unstable areas that lead to runaway evolution.  However, all of the joint CSSs we were able to generate always occurred in a region of low resilience; generally with less than $25\%$ of the potential maximum resilience (Figure \ref{coevcss}b; see appendix for extended treatment).  If the consumer and producer isoclines do not intersect, evolution proceeds via runaway selection along the consumer isocline and will drive traits $(s_P,s_H)$ to a minimum boundary if one is specified.  Finally, if the consumer isocline does not exist, evolution is sensitive to initial conditions and producer and consumer trait values either diverge toward infinity (or a specified bound) or proceed along the stable (right) part of the producer isocline where an ESS can exist for the producer but there is still runaway selection on $s_H$ for decreasing values \citep{loeuille2002}.  

\section{Discussion}

We have shown that an inherent conflict often exists between whole ecosystem resilience and component-wise evolutionary stability for a generic and widely used ecological model.  In particular, this is an example in which evolutionary stability at the producer and consumer levels results in low resilience of the whole ecosystem. When a producer trait that dictates nutrient uptake and influences grazing rate is allowed to evolve in a nutrient-producer-consumer ecosystem model, the trait value associated with the producer CSS does not confer the greatest resilience of the entire ecosystem.  Similarly, the evolution of a consumer trait does not confer the greatest resilience and it appears that coevolution of both producers and consumers typically does not lead to high-levels of resilience.
Using a similar model to the one analyzed here, \citet{loeuille2002} showed that trait evolution can induce dynamic instability (a transition from a stable node to a limit cycle), but we have gone further by establishing that it is often impossible to simultaneously maximize evolutionary stability and ecosystem resilience.  Together with the work of \citet{wilson1996}, \citet{loeuille2010}, and \citet{Loreau2010}, our results underscore the fact that evolution of species-level traits lead to less than maximally stable ecosystems.  This fact may force us to reevaluate our notions about ecosystem structure, in particular how it may emerge and be maintained.

Species-level trait evolution not only influences stability properties of entire ecosystems, but also material flux and the distribution of biomass.  For example, in the case of producer evolution only, the producer CSS does not maximize resource uptake, but rather strikes a balance between acquiring resources and avoiding predation, a commonly hypothesized trade-off (e.g. \citet{armstrong1979, Grover1994, leibold1996}), that has been experimentally verified more recently \citep{yoshida2004}.  In fact, the producer CSS in Figure \ref{contourer} is linearly related to consumer strategy across a gradient in consumer strategy, $s_H$.  A linear relationship between the producer strategy and the consumer strategy across a gradient in consumer strategy also appears for different values of the parameter $\chi$, which, controls the strength of the grazing interaction function $a(s_P,s_H)$.  As the consumer trait value increases, so should the producer trait to maintain evolutionary stability (Figure \ref{contourer}a).  In this way, the producer maintains a low level of consumption from the consumer across changes in the consumer strategy (Appendix Figure \ref{chiplotcss}).

We also found that the resilience of the ecosystem at the CSS is very low relative to the maximum resilience that could be achieved in general.  
For example, in the case of producer evolution only, there is a large quantitative difference between trait values associated with the producer CSS versus those associated with maximum resilience everywhere along a gradient of the consumer trait (Figure \ref{contourer}a).  The associated return time of the producer CSS is on the order of 20 units of time (e.g. days) compared to a return time of two units of time for the region of maximum resilience.  We found a similarly large difference between maximum resilience and resilience at the coevolutionary CSS.  In addition, maximum resilience does not always change smoothly with changes in trait values.  Maximum resilience and the producer trait ($s_P$) associated with maximum resilience change abruptly in some regions of trait space due to the two high resilience veins in Figure \ref{contourer}a.  Thus, at a specific value of the consumer trait ($s_H$), two different $s_P$ values have the same maximum resilience.  Hence, if one seeks to maximize resilience, very small differences in trait values can make the maximum attainable value change abruptly and to maintain maximum resilience, the ``lagging" trait would have to change significantly.  These multimodal peaks in resilience create the potential for nonlinear responses of an evolving ecosystem, the end result of which may only be a local maximum.  

\begin{figure}[h!]  \leavevmode \includegraphics[scale=0.7]
	{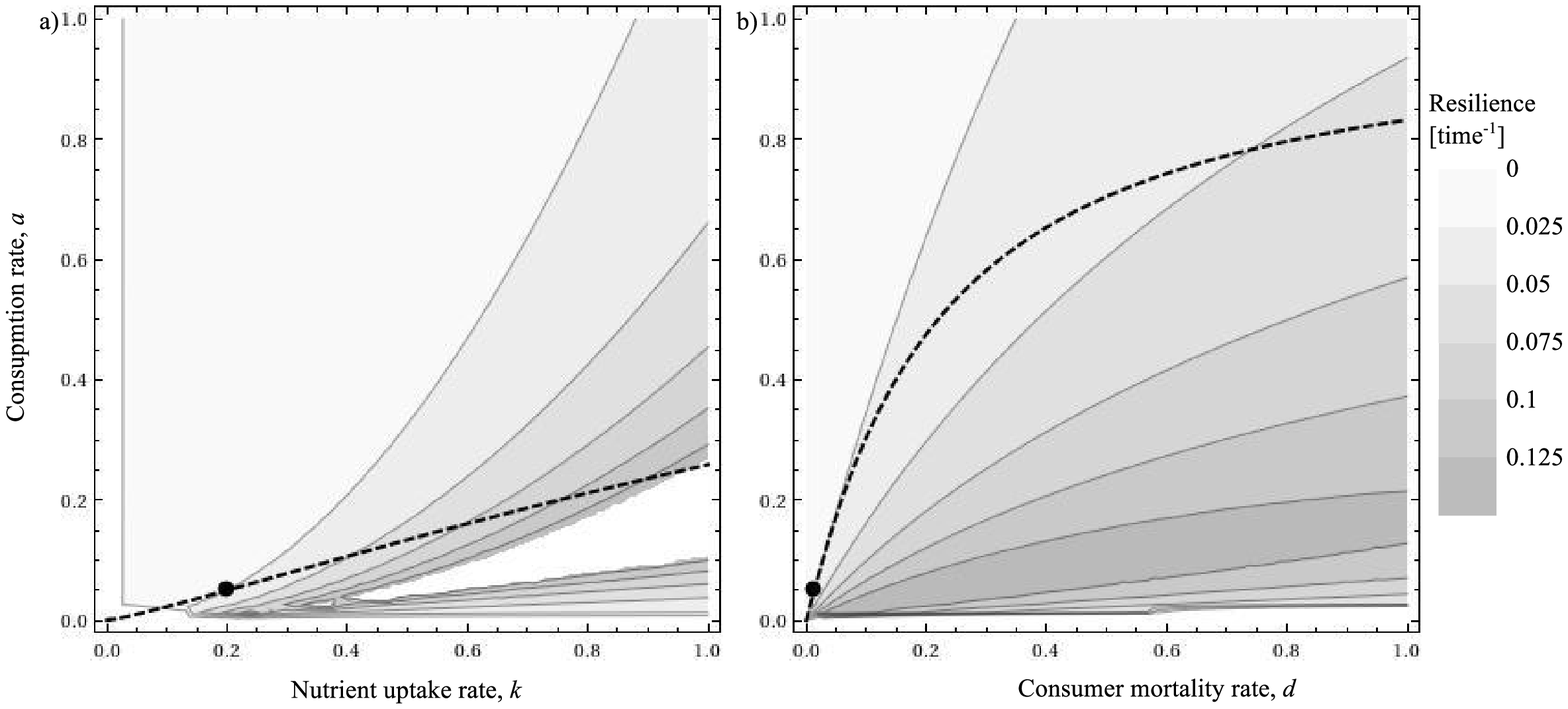}
		 \caption{Contour plot of resilience as a function of two traits, in a) $k$ (x-axis) and $a$ (y-axis) are the two traits and for b) $d$ (x-axis) and $a$ (y-axis) are the two traits.
	 Dashed line is tradeoff function used in figures in main text.  Black dot is the location of the singular strategy for coevolution, which, in this case is a CSS for both species (joint CSS).  Same parameters as Figure \ref{coevcss}b ($\chi=1$) except we use the trait values from the singular strategy to calculate: a) $d$ and for b) $k$.  Note that the resilience landscape also depends on these variables.  The white area in the interior of plot a) is for higher values of resilience than the top value of the scale shown.
 \label{tradeoffchi1}  
	} 
	\end{figure}

\begin{figure}[h!]  \leavevmode \includegraphics[scale=0.7]
	{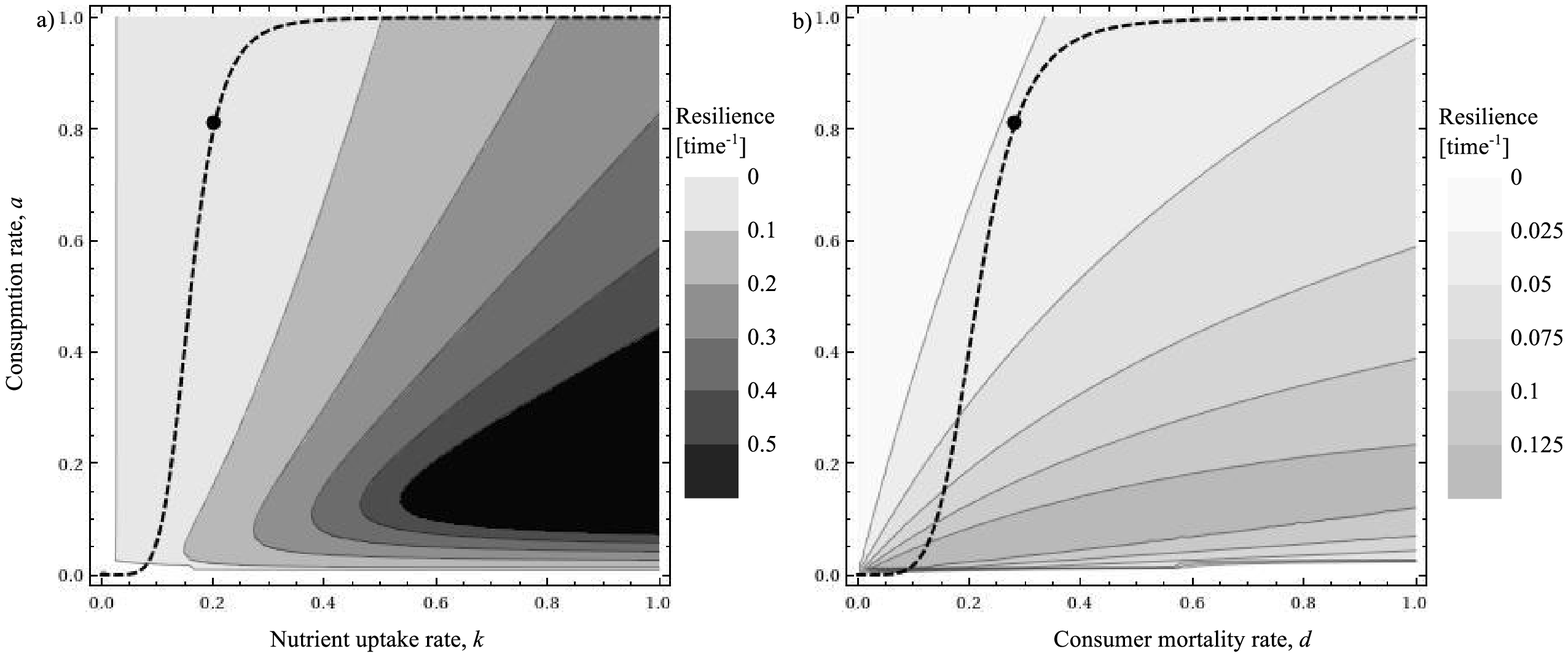}
		 \caption{Contour plot of resilience as a function of two traits, in a) $k$ (x-axis) and $a$ (y-axis) are the two traits and for b) $d$ (x-axis) and $a$ (y-axis) are the two traits.
	  Dashed line is tradeoff function used in figures in main text.  Black dot is the location of the singular strategy for coevolution, which, in this case is a CSS for the consumer only.  Same parameters as Figure \ref{coevcss}a ($\chi=5$) except we use the trait values from the singular strategy to calculate: a) $d$ and for b) $k$.  Note that the resilience landscape also depends on these variables.
 \label{tradeoffchi5}  
	} 
	\end{figure}

We used a previously employed tradeoff in our work (similar to \citet{loeuille2002, branco2010}), and our choice of the tradeoff shape may influence our results concerning stability.  Although we have not proved that maximum resilience and a CSS cannot coincide, we observed no correspondence between evolutionary stable outcomes and resilience over a wide range of parameters for the different tradeoff functions employed here (see Appendix Figures \ref{gridproev}, \ref{gridconev}, \ref{gridcoev}, \ref{tradeoffchi1} and \ref{tradeoffchi5}).  In general, with isolated producer evolution, the nature of the tradeoffs considered will place the producer CSS in a region of low resilience.  However, other outcomes are possible for isolated consumer evolution and coevolution.  In these cases, the shape of the tradeoff curve may permit the CSS to occur in a region of higher resilience by changing the location of the CSS and the resilience landscape.  We see this with varying $\chi$ (compare Figure \ref{tradeoffchi1} with Figure \ref{tradeoffchi5}).  With our tradeoffs and by comparing the geometry of potentially different tradeoffs in our model \citep{demazancourt2004}, we still only found CSSs, evolutionary repellers, and runaway selection, and therefore branching points are extremely unlikely, similar to the findings of \citet{branco2010}.

We were interested in characterizing how different combinations of resource acquisition quantitatively determine whole ecosystem resilience, and in particular, how resilience is related to $\Omega$.  
Other studies have attempted to link aggregate ecosystem measures, individual traits, and resilience and have found some interrelationships.  \citet{cropp2002} examined the five physiological parameters in their model food chain and used a genetic algorithm to find the maximum resilience as a function of those parameters.  They conclude that resilience is very strongly positively related to (producer max growth rate)/(grazing rate).  Although their model is different than ours, including their Type II producer-nutrient response, our $k(s_P)$ is essentially the same as their producer max growth rate.  Contrary to their findings, we observed a unimodal relationship of resilience with $\Omega=k(s_P)/a(s_P,s_H)$ (Figure \ref{ka}).  

Here, we have focused on one aspect of ecosystem stability, resilience, but other aspects of stability can and should be addressed in an evolutionary context in the future.  Resistance  \citep{Pimm1984} can play a more significant role in overall stability than resilience in model ecosystems \citep{Vallina2011}, future studies should consider the combined role of resistance and resilience in determining stability and the consequences of evolution since data suggest they are independent of one another (\citeauthor{steiner2006} \citeyear{steiner2006}, Mellard unpublished results).  In future studies, resistance should be calculated from a perturbation to individual system components so that it is a measure of whole system response to a component-level perturbation.  In addition, expanding the scope to include initially unstable systems, or those that cycle \citep{cortez2010} will enable a deeper understanding of the influence of evolution on the persistence of a population or ecosystem \citep{Allesina2012}.   Methods exist for characterizing stability in nonequilbrium environments \citep{neubert1997,Vallina2011}, and focus on the same basic idea of characterizing a recovery following a perturbation. 

Future studies should also explicitly study the diversity in or simultaneous evolution of traits associated with different trophic levels \citep{ellner2011} to fully understand the consequences of evolution for whole ecosystem resilience.  Future studies should also explicitly consider the degree to which phenotypic change matters \citep{ellner2011a}.  For example, behavioral responses may differ from species replacement, which may differ from evolution in how they affect whole-ecosystem properties and their responses to perturbations.  Expanding the focus to stability of the evolutionary equilibrium of different ecosystem models with richer dynamics \citep{abrams1996} could lead to a better understanding of the multiple layers of stability that affect whole ecosystem stability on different time scales and organizational levels.  

In closing, we reiterate that an essential tension can arise between evolutionary and ecological dynamics.  Using conventional ecological selection pressures at the population level, our results show that ecosystems tend to evolve to regions in trait space with relatively low overall resilience. The ecosystem model we studied here apparently cannot simultaneously maximize dynamical and evolutionary stability, as selection for each leads to very different trait values.  However, it remains to be seen if the discrepancy between traits values that confer maximum resilience and trait values corresponding to CSSs is a general feature of ecological systems.  Regardless, the fact that evolutionary stability and dynamical stability can be at odds has important implications for how we view and manage ecosystem structure.  Our results suggest that managing for biomass or production of individual species or ecosystem components may place the whole ecosystem at an increased risk of collapse.  

\section{Acknowledgements}
The authors would like to acknowledge several anonymous reviewers for providing helpful comments that led to improvements of the manuscript.  The Centre for Biodiversity Theory and Modelling is supported by the TULIP Laboratory of Excellence (ANR-10-LABX-41).

%%%%%%%%%%%%%%%%%%
\clearpage
\bibliographystyle{spbasic}
\bibliography{/Users/jarad/Dropbox/references}
%\begin{thebibliography}{50}
%\expandafter\ifx\csname natexlab\endcsname\relax\def\natexlab#1{#1}\fi

%\end{thebibliography}
%\bibliography{/Users/jaradmellard/Dropbox/references}
%%%%%%%%%%%%%%%%%%
\clearpage
\newpage
\section{Appendix}
%{\bf Appendix S1}
\subsection{Stability of ecological equilibrium}
We classify the stability of the interior equilibrium.  The elements of the Jacobian matrix at equilibrium (\ref{jacobian}) form the characteristic equation
\begin{equation}
a_3 \lambda^3+a_2 \lambda^2+a_1 \lambda+a_0=0. %\nolabel
\end{equation} 
In order to be a stable equilibrium, the coefficients must satisfy the following Routh-Hurwitz criteria \citep{may:1973}: $a_n>0$ and $a_2a_1>a_3a_0$.
It is easy to show that $a_3>0$ and $a_2>0$.  For the other coefficients, $a_1>0$ if $\frac{I \Omega(k l + b)}{q + \Omega \Psi} > m b$ and $a_0>0$ if $I \Omega > m(q + \Omega \Psi)$.  The second Routh-Hurwitz criterion, $a_2a_1>a_3a_0$, can be simplified to $k l > 0$, which is always true since both of these parameters are always $(+)$.

\clearpage
\newpage
\subsection{Influence of other parameters}

%\clearpage
\begin{figure}[h!]  \leavevmode \includegraphics[scale=0.7]
	{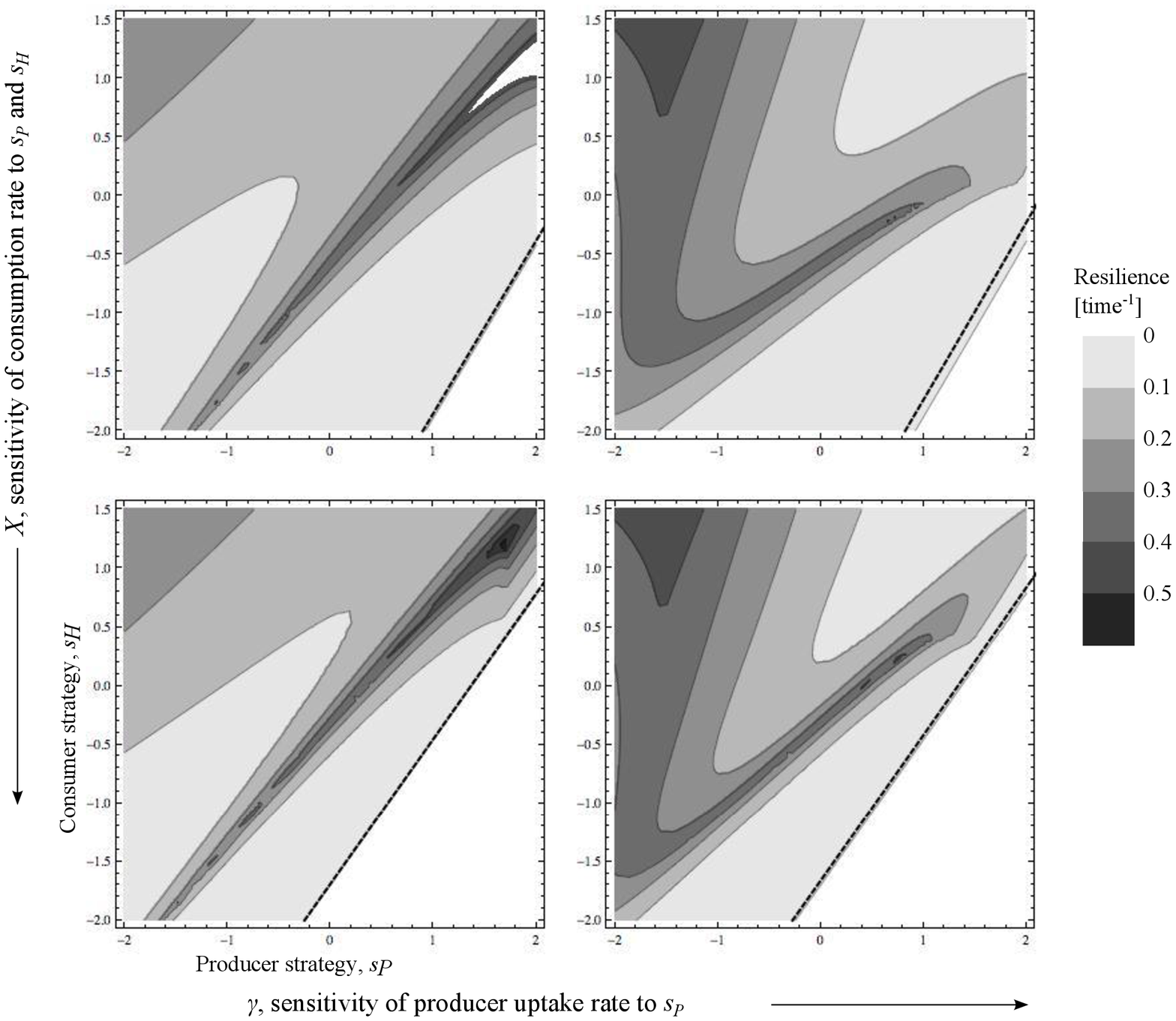}
		 \caption{Contour plot of resilience as a function of two traits producer strategy, $s_P$ and consumer strategy, $s_H$ arrayed across several other parameters, in this case a parameter describing the steepness of the species interaction function to the traits, $\chi$ on the vertical axis, and a parameter describing the steepness of the producer nutrient uptake rate to its trait, $\gamma$ on the horizontal axis.  Darker regions are higher levels of resilience with the exception that the white region in the interior of the top left plot is for higher values of resilience than the top value.  The lower right hand corner white region is a feasibility boundary where consumers cannot exist in the system ($\hat{H}=\frac{I l}{\frac{q}{\Omega}+\Psi}-\frac{m}{a} \leq 0$).  Resilience calculated according to equation \ref{resilience}.  In this scenario, we assume producer evolution only and the trait values that comprise the producer CSS are denoted by the small dashed line on the plot.  Parameters are:  $\alpha=1, \beta=1, d_H=0.05, \omega=0.75, g=0.95$.   The parameter $\gamma$ ranges from 0.24 to 0.74 in units of .5 and the parameter $\chi$ ranges from 3-5 in units of 2. 
 \label{gridproev}  
	}
	\end{figure}

\clearpage
\begin{figure}[h!]  \leavevmode \includegraphics[scale=0.7]
	{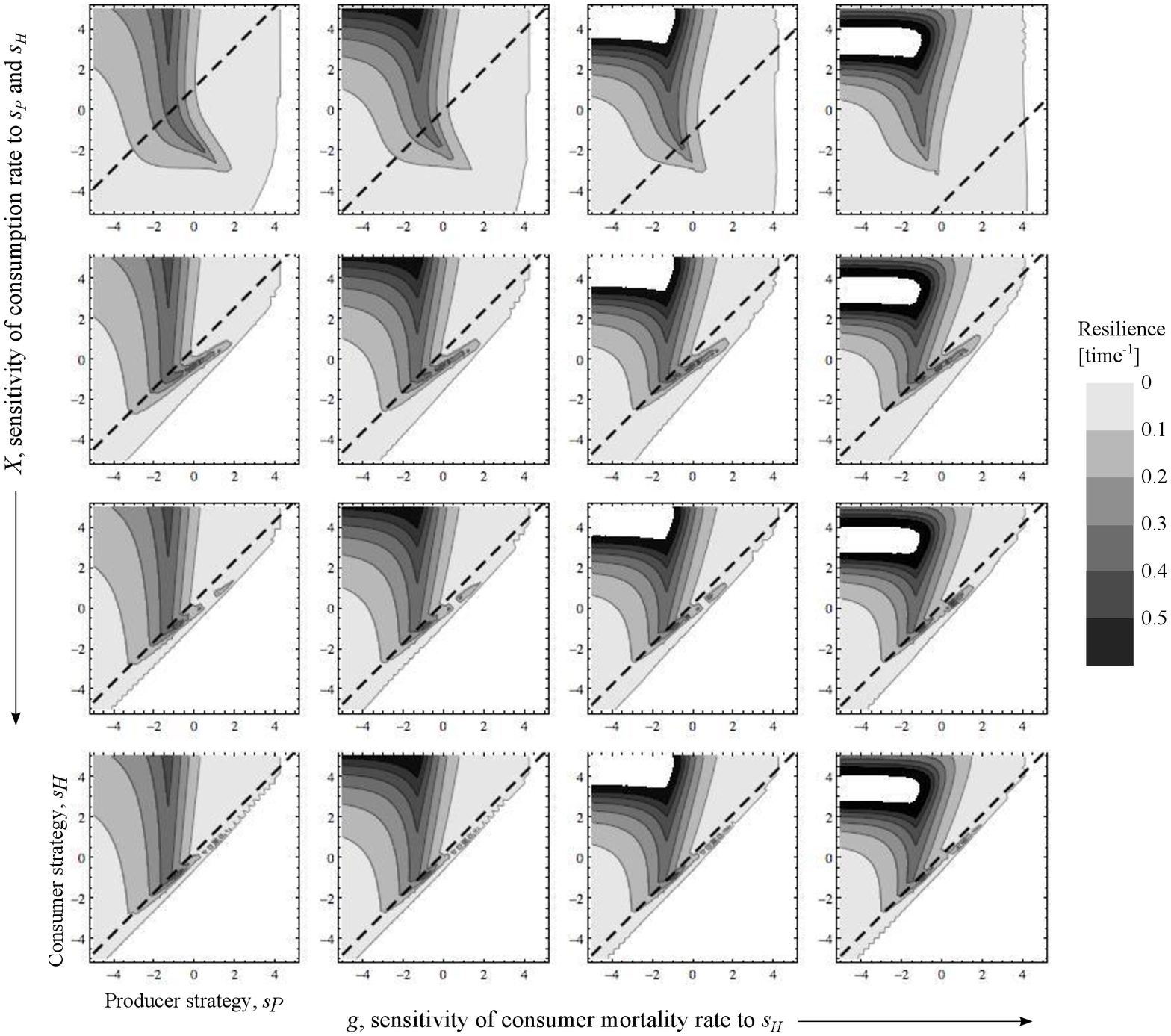}
		 \caption{Contour plot of resilience as a function of two traits producer strategy, $s_P$ and consumer strategy, $s_H$ arrayed across several other parameters, in this case a parameter describing the steepness of the species interaction function to the traits, $\chi$ on the vertical axis, and a parameter describing the steepness of the consumer mortality rate to its trait $g$ on the horizontal axis.  Darker regions are higher levels of resilience with the exception that the white region in the upper left of the plots is for higher values of resilience than the top value.  The lower right hand corner white region is a feasibility boundary where consumers cannot exist in the system ($\hat{H}=\frac{I l}{\frac{q}{\Omega}+\Psi}-\frac{m}{a} \leq 0$).  Resilience calculated according to equation \ref{resilience}.  In this scenario, we assume consumer evolution only and the trait values that comprise the consumer CSS are denoted by the large dashed line on the plot.  Parameters are:  $\alpha=1, \beta=1, d_H=0.05, \omega=0.75, \gamma=0.85$.   The parameter $g$ ranges from 0.24 to 0.99 in units of .25 and the parameter $\chi$ ranges from 1-13 in units of 4.	
 \label{gridconev}  
	}
	\end{figure}

\clearpage
\begin{figure}[h!]  \leavevmode \includegraphics[scale=0.7]
	{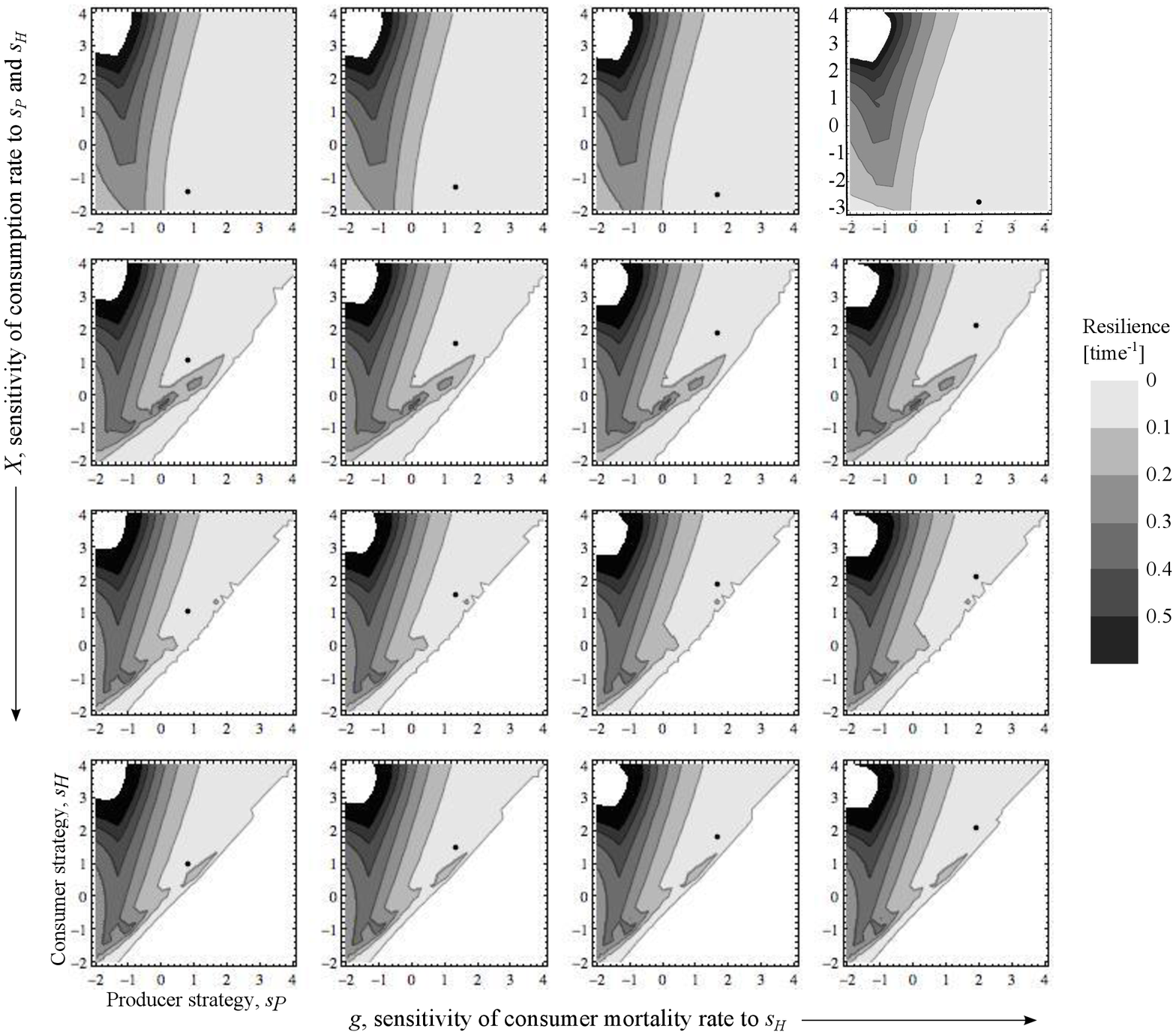}
		 \caption{Contour plot of resilience as a function of two traits producer strategy, $s_P$ and consumer strategy, $s_H$ arrayed across several other parameters, in this case a parameter describing the steepness of the species interaction function to the traits, $\chi$ on the vertical axis, and a parameter describing the steepness of the consumer mortality rate to its trait, $g$ on the horizontal axix.  Darker regions are higher levels of resilience with the exception that the white region in the upper left of the plots is for higher values of resilience than the top value of the scale shown.  The lower right hand corner white region is a feasibility boundary where consumers cannot exist in the system ($\hat{H}=\frac{I l}{\frac{q}{\Omega}+\Psi}-\frac{m}{a} \leq 0$).  Resilience calculated according to equation \ref{resilience}.  In this scenario, we assume coevolution and the black dot is the location of the joint CSS.  Parameters are:  $\alpha=1, \beta=1, d_H=0.05, \omega=0.75, \gamma=0.85$.   The parameter $g$ ranges from 0.9 to 0.99 in units of .03 and the parameter $\chi$ ranges from 1-13 in units of 4.		 
 \label{gridcoev}  
	} 
	\end{figure}

\subsection{Interpreting the CSS}
\subsubsection{Example:  Location of producer CSS with producer evolution only}
\begin{figure}[h!] 
\centering \leavevmode \includegraphics[scale=0.85]
	{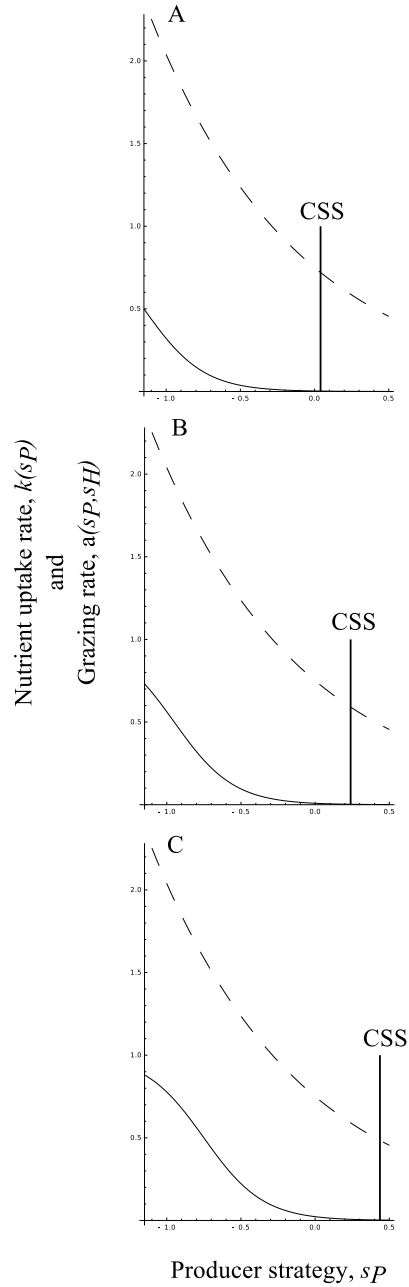}
	 \caption{
	Location of the producer CSS (when only the producer is evolving) on the nutrient uptake  [(producer biomass x time)$^{-1}$] and grazing [(consumer biomass x time)$^{-1}$] function curves for different values of $s_H$:  A) $s_H=-1.15$, B) $s_H=-0.95$, C) $s_H=-0.75$.  Same curves as in Figure \ref{chiplotsallgrazweak}.  Vertical black line points to the location of the producer CSS on the $s_P$ axis.  Dashed line is a plot of  $k(s_P)$ where $\omega=0.75$.  Solid lines are plots of $a(s_P,s_H)$ where $\alpha=1, \beta=1, \chi=5$.
\label{chiplotcss}  
	} 
	 \end{figure}

The CSS $s_P$ of the producer is partly determined by the consumer strategy $s_H$ because the consumer trait $s_H$ determines which $s_P$ strategies get grazed on most heavily.  In this model, the producers are limited by grazing, which means that although there is a tradeoff between nutrient uptake and susceptibility to grazing, the producer CSS strategy mainly conforms to what strategy reduces grazing the most (while keeping nutrient uptake at as high of levels as possible).  For example, in Figure \ref{chiplotcss}, the producer CSS strategies are all located on the tails of the grazing versus $s_P$ curve.  Note that  no consumers in the system leads to runaway selection for maximum nutrient uptake of the producers.

\clearpage
\newpage
\subsection{Tradeoff shape}
%\clearpage
	
\end{document}